\documentclass[]{spie}
\usepackage[]{graphicx}
\usepackage[]{astro}

\def\microas{\mu{\rm as}}

\title{Scientific Results from High-precision Astrometry at
the Palomar Testbed Interferometer}

\author{Matthew W.~Muterspaugh\supit{a}
Benjamin F.~Lane\supit{b} 
Maciej Konacki\supit{c}
B.~F.~Burke\supit{b}
M.~M.~Colavita\supit{d}
S.~R.~Kulkarni\supit{e}
M.~Shao\supit{d}
\skiplinehalf
\supit{a}Department of Geological and Planetary Sciences, California 
Institute of Technology, Pasadena, CA 91125; \\
\supit{b}Massachusetts Institute of Technology, Kavli Institute for 
Astrophysics and Space Research, Department of Physics, 70 Vassar Street, 
Cambridge, MA 02139; \\
\supit{c}Nicolaus Copernicus Astronomical Center, 
Polish Academy of Sciences, 
Rabianska 8, 87-100 Torun, 
Poland; \\
\supit{d}Jet Propulsion Laboratory, California Institute of Technology, 
4800 Oak Grove Dr., Pasadena, CA 91109; \\
\supit{e}Division of Physics, Mathematics and Astronomy, 105-24, California 
Institute of Technology, Pasadena, CA 91125
}

\authorinfo{\noindent Send correspondence to M.W.M.\\
M.W.M.: E-mail: matthew1@ssl.berkeley.edu, Telephone: 1 617 452 2304
}

\begin{document} 
\maketitle

\noindent
{\bf Copyright 2006 Society of Photo-Optical Instrumentation Engineers.\\}
This paper will be published in SPIE conference proceedings volume 6268, 
``Advances in Stellar Interferometry.''  and is made available as 
and electronic preprint with permission of SPIE.  One print or electronic 
copy may be made for personal use only.  Systematic or multiple reproduction, 
distribution to multiple locations via electronic or other means, duplication 
of any material in this paper for a fee or for commercial purposes, or 
modification of the content of the paper are prohibited.

\begin{abstract}
A new observing mode for the Palomar Testbed Interferometer was developed in
2002-2003 which enables differential astrometry at the level of 20
micro-arcseconds ($\microas$) for binary systems 
with separations of several hundred
milli-arcseconds (mas). This phase-referenced mode is the basis of the Palomar
High-precision Astrometric Search for Exoplanet Systems (PHASES), a search
for giant planets orbiting either the primary or secondary star in fifty
binary systems. We present the first science results from the PHASES
search. The properties of the stars comprising binary systems are
determined to high precision. The mutual inclinations of several
hierarchical triple star systems have been determined. We will present
upper limits constraining the the existence of giant planets in a few of
the target systems.
\end{abstract}


\section{Introduction}
\label{sect:intro}  

The Palomar High-precision Astrometric Search for Exoplanet Systems (PHASES) 
uses phase-referencing techniques to perform differential astrometry of 
sub-arcsecond binaries at the level of 20 $\microas$.
These observations provide precision visual orbits of the binaries and 
allow detection of tertiary components orbiting either the primary or 
secondary due to the reflex motion of the subsystem center-of-light.  
A detailed description of the observational method and data analysis 
procedures was presented by Lane \& Muterspaugh\cite{LaneMute2004a}.

PHASES data are collected at the Palomar Testbed Interferometer 
(PTI) \cite{col99}, located on Palomar Mountain near San Diego,
CA. It was developed by the Jet Propulsion Laboratory,
California Institute of Technology for NASA, as a testbed for
interferometric techniques applicable to the Keck Interferometer and
other missions such as the Space Interferometry Mission (SIM).  It
operates in the J ($1.2 \mu{\rm 
m}$), H ($1.6 \mu{\rm m}$), and K
($2.2 \mu{\rm m}$) bands, and combines starlight from two out of three
available 40-cm apertures. The apertures form a triangle with one 110
and two 87 meter baselines.

The primary goal of the PHASES program is to find and characterize 
giant planets in close binary systems.  The existence of such systems 
poses strong challenges for models of giant planet formation.  While 
it is possible each of the two processes currently favored---core accretion 
\cite{Liss1993} and gravitational instability \cite{Boss2000}---contribute 
to giant planet formation around single stars and wide binaries, simulations 
show both schemes have obstacles when a second star orbits so closely that it 
interacts with the planet-forming circumstellar 
disk \cite{Nelson2000}, truncating it in size and heating it.  In 
four exoplanet hosting binaries---HD 188753 \cite{Konacki2005}, 
$\gamma$ Cephei \cite{Hatzes2003}, GJ 86 \cite{Queloz2000}, 
and HD 41004 \cite{Zuc2004}---the secondary star 
would have truncated the disks to less than 6 AU.  It is possible 
that some of these systems reached their current configurations via dynamical 
interactions in the short-lived star clusters in which they formed 
\cite{Pfahl2005}\cite{Portegies_Zwart_2005} , 
though this post-formation mechanism appears too 
infrequent to explain the number found.  These systems have been identified 
with the radial velocity (RV) method.  PHASES employs astrometry, from which 
the companion mass can be identified without the ambiguity of the orbital 
inclination.  Furthermore, the relative orientations of the binary and 
planetary orbits can be determined in order 
that system dynamics can be studied.  
This effort and others specifically targeting close binaries will better 
determine the frequency of planets in binaries; if large, this will be strong 
motivation for a revolution in the theory of giant planet formation.

\section{Method}

The challenge of astrometric detections of planets in binaries 
comes from the small size of the reflex motions they cause 
to the host stars.  Giant planets mass a factor of $1000$ less 
than stars, and are only stable in orbits sized of order 1/10 of 
the binary separation or smaller.  Thus, one requires an 
astrometric precision of order $10^{-4}$ of the binary separation.
This goal drives one to use high spatial resolution techniques.

An optical interferometer coherently combines light from two or 
more telescopes to measure the resulting sinusoidal intensity 
variations that result from constructive and destructive 
interference of the starlight.  These ``fringes'' form the basic 
observable for interferometric imaging.  Constructive combination requires 
that the total optical paths from a star through each of the telescopes 
are matched to identical wavefront phases.  The optical bandpass 
creates an envelope function modulating the fringe contrast, 
resulting in an observable fringe packet.  For systems with finite 
optical bandpasses, fringes with large contrast form only within a 
limited range of differential optical paths, centered at a zero value where 
the optical pathlengths are identical for the interferometer arms being 
combined.

The optical path lengths traversed by the starlight have several terms.  
First, variations in atmospheric indices of refraction over each telescope 
can introduce time-variable differential path terms.  
Second, if a target is not directly overhead, one telescope will be slightly 
closer than the other, introducing a geometric term $\vec{B}\cdotp\vec{s}$ 
where $\vec{B}$ is the baseline vector connecting two telescopes, and 
$\vec{s}$ is the unit vector in the direction of the star.  Finally, 
path variations internal to the interferometer must be included; in practice, 
these internal paths are actively varied to compensate for the first two 
terms.

\begin{figure}[htbp]
   \begin{center}
   \begin{tabular}{c}
   \includegraphics[height=3.5in]{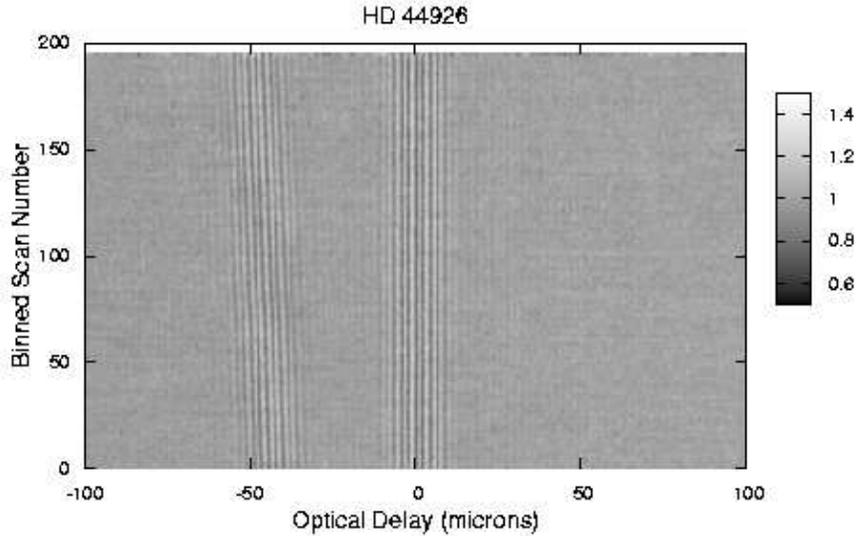}
   \end{tabular}
   \end{center}
   \caption[Sample Interferograms]
   {\label{fig:waterfall} Measured intensity in the detector as a
function of differential optical path, for successive scans of the
speckle binary system HD 44926. Each scan takes 1.5 seconds to
acquire. The fringe tracker was locked on to the bright star (around
0), while the second star produces a fringe pattern which starts at -40
$\mu$m and moves due to Earth rotation. Although the second
fringe pattern is relatively faint, the effect of coherently 
co-adding 500--2000 scans produces a high signal-to-noise 
ratio in the final astrometric measurement.}
\end{figure}

Note that the geometric term $\vec{B}\cdotp\vec{s}$ varies with 
sky position.  A binary star will produce two 
fringe packets centered at slightly different delays.  For PHASES targets, 
the packets do not overlap, 
see figure \ref{fig:waterfall}.  If their separations 
are small, the atmospheric contribution to the differential path length 
is correlated spatially.  Thus, measuring the differences in internal delays 
required to produce each fringe packet 
determines the binary separation projected along the baseline vector.

While the atmospheric terms are correlated spatially, they vary in time.  
These variations must be monitored if the fringe packets are measured at 
different times.  The monitoring and correcting of these fluctuations 
are known as ``phase-referencing''\cite{lc03}.  
PTI's original phase-referencing mode operated on two sources 
separately resolvable by the interferometer telescopes.  This is modified to 
a simpler version that studies 
all components within one telescope resolution element.  Two interferometric 
beam combiners are used: one tracks just a single 
fringe of any star in the one-arcsecond 
field and operates on fast (10 ms) timescales to monitor the atmosphere, 
the other one makes precise measurements of the relative positions 
of each star in the atmosphere-stabilized field.  This latter 
system samples both fringe packets by introducing a triangle-wave 
internal path modulation with amplitude $\sim 100-300 \mu {\rm m}$ and 
period $\sim 1-3$ seconds.  Several thousand scans through the double 
interferograms are recorded in an hour of observation.
In post-processing the optical path separation of 
the binaries is converted to projected
sky separation via an algorithm that depends on the baseline of the
interferometer and sky position of the target.

Night-to-night repeatability at the $\sim 20 \microas$ level has been 
demonstrated for many target systems with binary separations of order 
$\sim 200$ mas.  This matches the $10^{-4}$ relative precision needed to 
detect giant planets.

\section{Limits on Tertiary Companions}

PHASES observations of several binaries span 2-3 years.  It is now 
possible to search for perturbations to the Keplerian orbits that would 
indicate the presence of tertiary companions, and determine the mass 
threshold above which companions can be shown not to exist.  This mass 
threshold depends on the companion orbital period and these two parameters 
form a phase-space over which one can search for orbital configurations 
that are consistent with the observations.  This search is computationally 
intensive and is currently limited to face-on, circular orbits.

\subsection{$\delta$ Equulei}

\begin{figure}[h]
\centerline{\includegraphics[width = 5.0in]{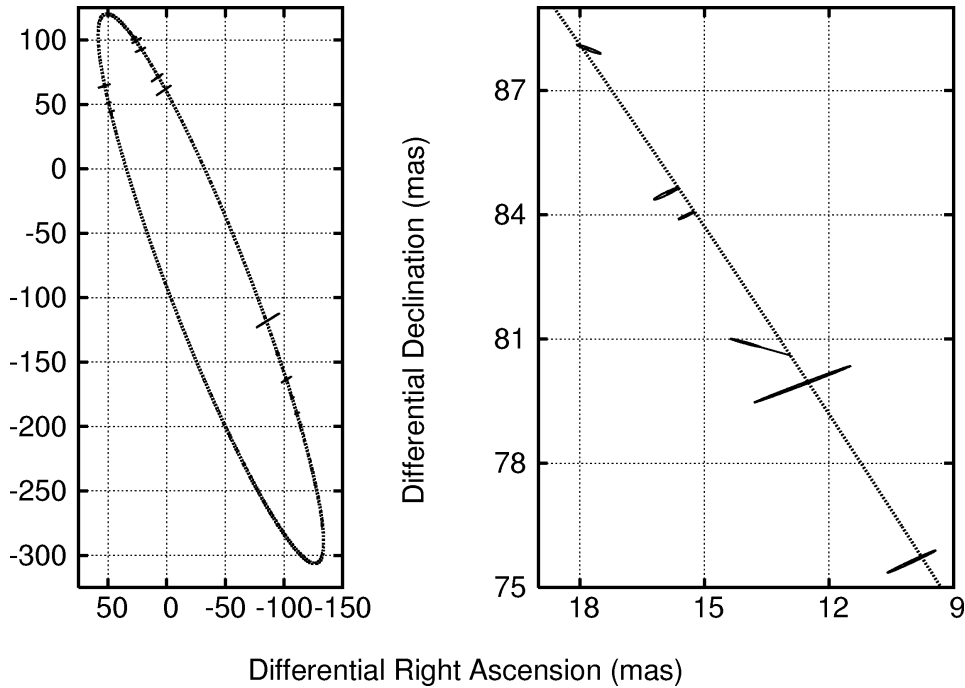}}
\caption[$\delta$ Equulei Visual Orbit] 
{ \label{fig:202275_orbit}
Visual orbit of $\delta$ Equulei with PHASES observations.  
}
\end{figure}

The combined visual and RV orbit of $\delta$ Equulei 
(7 Equ, HR 8123, HD 202275) was described by 
Muterspaugh et al.\cite{Mut05_delequ}  That study was able to determine the 
component masses at the $ 1 \% $ level and the distance to the system at a 
precision of 0.05 pc; improvement on these quantities will require 
additional RV observations.  In the 2005 observing season, an 
additional 11 PHASES measurements have been made, 
bringing the total to 38.  The new visual orbit is show in figure 
\ref{fig:202275_orbit} with the PHASES observations.

\begin{figure}[h]
\centerline{\includegraphics[width = 2.75in]{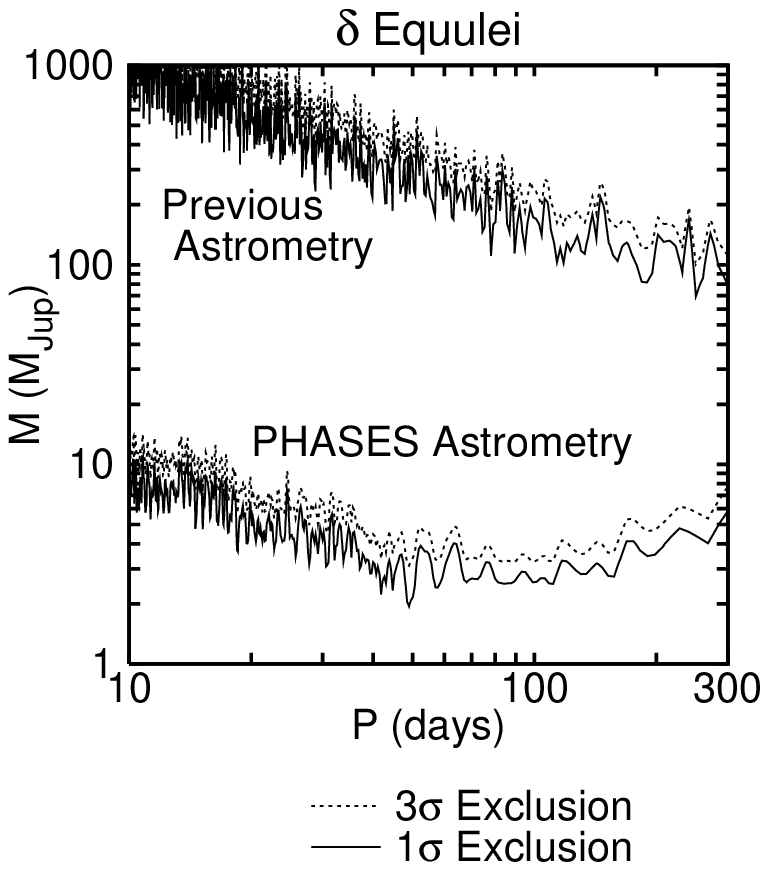}\includegraphics[width = 2.75in]{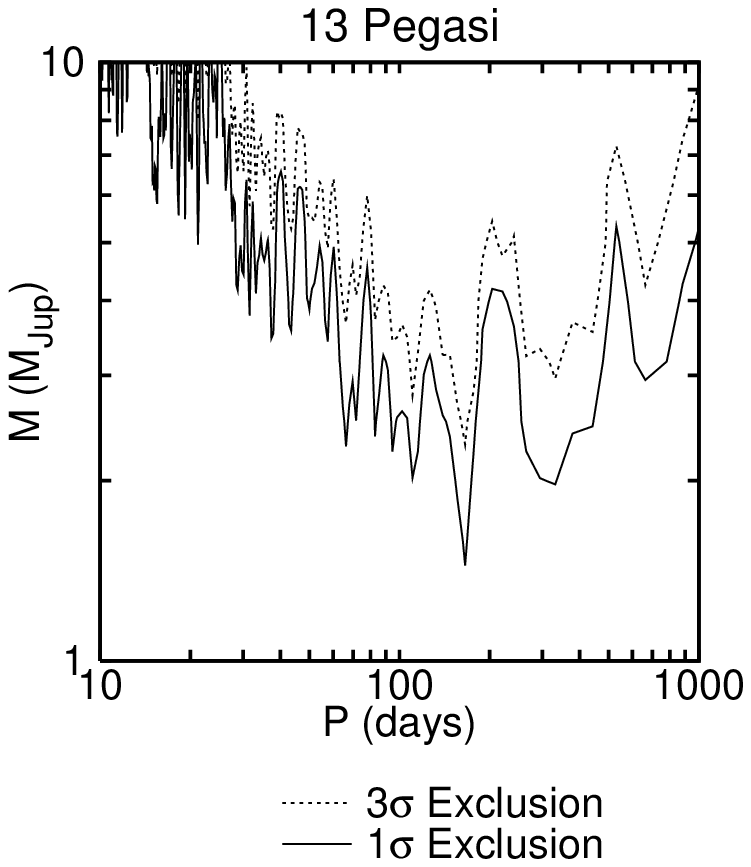}}
\caption[Mass-Period Companion Phase Space] 
{ \label{fig:phase_space}
The Mass-Period phase-space for tertiary companions that can be excluded by 
PHASES observations of two binary star systems.  On the left, $1-$ 
(solid line) and $3-\sigma$ (dashed line) $\chi^2$ contours for 
$\delta$ Equulei are shown; the corresponding contours based solely on 
micrometer and speckle interferometry data are also shown.  On the right, 
the same contours for 13 Pegasi, based only on PHASES measurements.  Note that 
a range of giant planets in face-on orbits are ruled out by the PHASES 
observations.
}
\end{figure}

Figure \ref{fig:phase_space} shows the regions 
in companion mass-period phase space for which the measurements are 
inconsistent with a perturbation caused by a hypothetical faint 
companion in a face-on circular orbit.  Companions with masses 
greater than the line plotted are inconsistent with the PHASES 
observations; the excluded regions are those above the lines.
Also shown are the exclusion regions for 170 differential astrometry 
measurements made by visual micrometry and speckle interferometry.  
The stability criteria of Holman \& Wiegert \cite{holman1999} indicates 
that planetary companions are stable only in orbits of $2/3$ year or less.

\subsection{13 Pegasi}

Twenty-five PHASES measurements have been made of 13 Pegasi 
(HR 8344, HD 207652).  Massive planets 
in 20 day to 3 year period face-on circular orbits would perturb this binary 
by more than the observed scatter in the PHASES data.  Planets as small as 
two Jupiter masses are ruled out in $\sim 4$ month period orbits.

\section{Triple Star Systems}

Studies of binary systems combining visual and RV orbits are able to 
measure component masses and the distance to the system (without 
relying on parallax observations), quantities that cannot be 
extracted from observations using only one method or the other.  
A star's mass is the most important property determining its 
evolution.  Measuring the distances to binaries can be a crucial 
step in establishing astrophysical distance scales, as in the case of 
the Pleiades \cite{Pan2004}\cite{npoiAtlas} .

Until recently the number of binary systems for which 
such combined visual and RV 
orbits were possible was severely limited by the conflicting 
observational biases for visual and RV measurements; visual orbits are 
more easily measured for long-period, widely separated systems, whereas 
velocities are largest in short-period systems.  
The gap between visual and RV measurements has been bridged by the 
maturation of long baseline optical interferometry, which has been used to 
obtain high resolution measurements of the visual orbits of spectroscopic 
binaries\cite{hummel98}\cite{boden_iota_peg} .  

The problem is compounded for hierarchical triple star systems (in which one 
component of the ``wide'' binary is itself a much shorter period ``narrow'' 
binary).  In order that the RV amplitude of the wide pair is large enough to 
be detected, its orbital period must be equally short (and have 
corresponding separation as small) as the two-component binaries in 
combined RV/visual studies.  As the ``narrow'' 
pair is necessarily smaller (by a factor of $\sim 10$ or more), resolving this 
orbit requires extending the RV/visual overlap by an additional order 
of magnitude.

To study the mutual inclination of the orbits in such a system, both RV and 
visual orbits for both wide and narrow 
systems in a triple are required.  Thus, 
determinations of the mutual inclinations of the two orbits in hierarchical 
triple stellar systems are rare, with only four previous systems having been 
studied.  Mutual inclination measurements 
are useful in studying properties of the 
environments in which stars form; the dynamical relaxation process undergone 
by multiples after formation is expected to leave a statistical 
``fingerprint'' in the distribution of mutual inclinations\cite{Sterzik2002}.
This provides motivation to overcome the 
challenge that extending the RV/visual 
overlap presents.  PHASES differential astrometry operating on subarcsecond 
binaries with a relative precision of 
$10^{-4}$ can identify the center-of-light 
perturbation due to the Keplerian orbit of a subsystem in one of the two 
components.

\begin{figure}[h]
   \centerline{\includegraphics[height=2.5in]{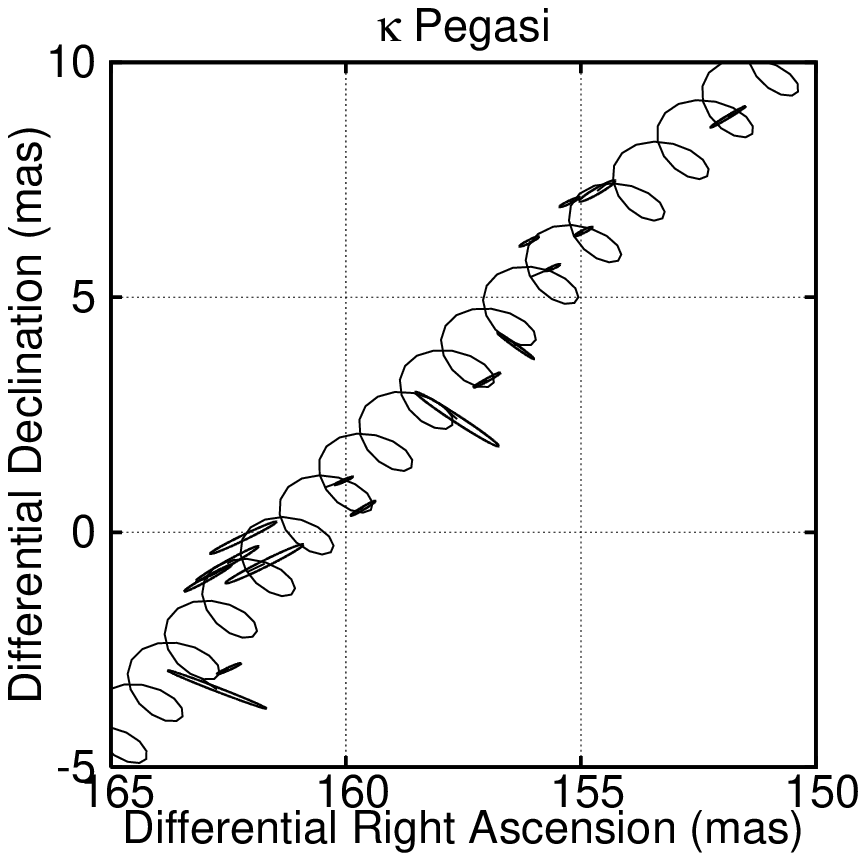}\hspace{0.25in}\includegraphics[height=2.5in]{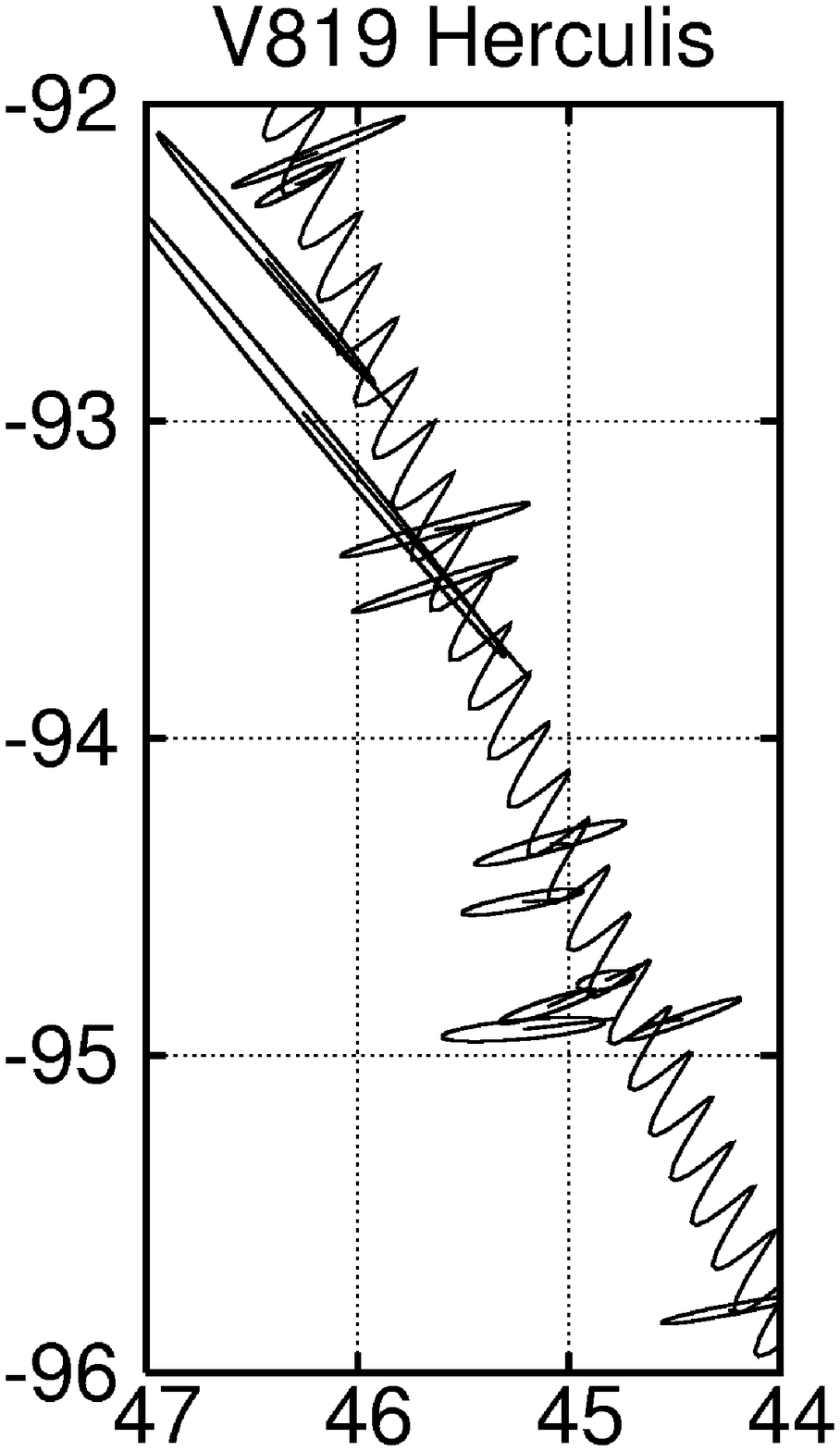}}
   \caption[Visual orbits of $\kappa$ Pegasi and V819 Herculis]
    { \label{tripleOrbits}
      The visual orbits of $\kappa$ Pegasi (left) and V819 Herculis (right) 
      showing perturbations by third components.
    }
\end{figure}

$\kappa$ Pegasi is a well-known, nearby triple star system.  It consists of 
a ``wide'' pair with semi-major axis 235 mas, one component 
of which is a single-line spectroscopic binary (semi-major axis 
2.5 mas); see figure \ref{tripleOrbits}.  
Using PHASES differential astrometry and iodine-cell 
referenced radial velocity observations, the masses for all three 
components are determined and the relative inclination between the wide and 
narrow pairs' orbits is found to be $43.8 \pm 3.0$ degrees, just over the 
threshold for the three body Kozai resonance \cite{Mut06_kappeg}.  
The system distance is 
determined to $34.64 \pm 0.22$ parsec, and is consistent with trigonometric 
parallax measurements.  The PHASES observations show conclusively that a 
previously suggested additional (fourth) 
stellar component is not present in the 
system.  Future investigations of this system to determine the relative 
luminosities of the three components will allow model fitting of the 
components' evolutions, of particular interest because two components have 
evolved slightly off the main sequence.

\begin{figure}[h]
   \centerline{\includegraphics[width=3.0in]{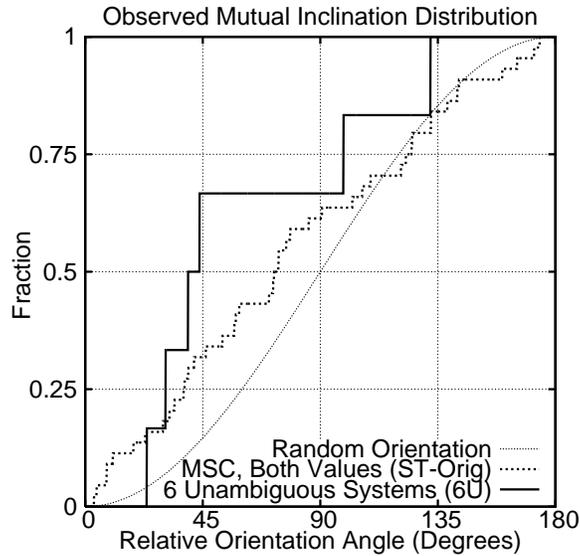}}
   \caption[Observed Angular Momentum Orientation Distribution]
    { \label{mutualInclinationCDFplot}
    Cumulative distribution of the six systems with unambiguous 
    mutual inclinations (set 6U).  The set ST is that of Sterzik \& Tokovinin 
    \cite{Sterzik2002}, who included both possible 
    mutual inclination angles for each of 22 systems for which 
    only ambiguous values were possible.  By including 
    both possible angles, distribution ST may 
    be ``diluted'' and partially biased 
    towards randomly oriented orbits.  The theoretical distribution for 
    random orientations is also shown.  The difference between 
    6U and random is statistically significant at the 
    96\% level.
    }
\end{figure}

Using PHASES observations, a similar study was made of the V819 Herculis 
system\cite{Mut06_v819her} (HR 6469, HD 157482).  The mutual inclination is 
found to be $23.6 \pm 4.9$ degrees.  It should be 
noted that the center-of-light wobble 
of the V819 Her narrow pair is $\sim 110 \microas$; this same amplitude would 
identify a planetary companion if the pair's orbital period 
were several months instead of two days.  These two triples 
represent just the fifth and sixth unambiguous determinations of the mutual 
inclinations of the orbits in hierarchical triple star systems.  
The distributions of the six known systems disagree (at the $ 96 \%$ level) 
with what one would expect from random orientations, suggesting structure 
in the angular momentum distributions of star forming regions; see figure 
\ref{mutualInclinationCDFplot}.

\section{Conclusions}

The PHASES program at PTI has been making routine observations for 2-3 years 
with relative astrometric precisions at the $10^{-4}$ level.  The first 
results have been high precision orbits of binary and triple systems, 
including measurements determining the mutual inclinations of the wide and 
narrow pairs of two triples.  Observations are starting to rule out the 
existence of planet-mass companions in several systems, and will be able to 
detect giant planets in longer period orbits as the search continues over 
the next years.  Detecting planetary companions in these close binaries 
will challenge current models of planet formation, and the astrometric nature 
of the observations will present uniquely detailed information on the 
companions, including mass without inclination ambiguities.

\acknowledgments

PHASES benefits from the efforts of the PTI collaboration members who have 
each contributed to the development of an extremely reliable observational 
instrument.  Without this outstanding engineering effort to produce a solid 
foundation, advanced phase-referencing techniques would not have been 
possible.  We thank PTI's night assistant Kevin Rykoski for his efforts to 
maintain PTI in excellent condition and operating PTI in phase-referencing 
mode every week.  Part of the work described in this paper was performed at 
the Jet Propulsion Laboratory under contract with the National Aeronautics 
and Space Administration. Interferometer data was obtained at the Palomar
Observatory using the NASA Palomar Testbed Interferometer, supported
by NASA contracts to the Jet Propulsion Laboratory.  This research has made 
use of the Washington Double Star Catalog maintained at the U.S.~Naval 
Observatory.  This research has made use of the Simbad database, operated 
at CDS, Strasbourg, France.  MWM acknowledges the MIT Earth, Atmospheric, 
and Planetary Sciences department for hosting him while this paper was 
written.  BFL acknowledges support from a Pappalardo Fellowship in Physics.  
PHASES is funded in part by the California Institute of Technology Astronomy 
Department, and by the National Aeronautics and Space Administration under 
Grant No.~NNG05GJ58G issued through the Terrestrial Planet Finder Foundation 
Science Program.  This work was supported in part by the National Science 
Foundation through grants AST 0300096 and AST 0507590.

\bibliography{muterspaugh_spie_2006}   
\bibliographystyle{spiebib}   

\end{document}